\documentclass[%
 aip,
 amsmath,amssymb,
 reprint,%
]{revtex4-1}

\usepackage{graphicx}
\usepackage{dcolumn}
\usepackage{bm}

\usepackage[utf8]{inputenc}
\usepackage[T1]{fontenc}
\usepackage{mathptmx}
\usepackage{etoolbox}
\usepackage{xcolor}

\makeatletter
\def\@email#1#2{%
 \endgroup
 \patchcmd{\titleblock@produce}
  {\frontmatter@RRAPformat}
  {\frontmatter@RRAPformat{\produce@RRAP{*#1\href{mailto:#2}{#2}}}\frontmatter@RRAPformat}
  {}{}
}%
\makeatother
\begin{document}

\preprint{AIP/123-QED}

\title{Fragmented charged domain wall below the tetragonal-orthorhombic phase transition in BaTiO$_3$}
\author{Petr S. Bednyakov}
\email{bednyakov@fzu.cz}
\altaffiliation{
FZU-Institute of Physics of the Czech Academy of Sciences$,$ Na Slovance 2$,$ 18221 Praha 8$,$ Czech Republic}

\author{Iegor Rafalovskyi}%
\affiliation{ 
FZU-Institute of Physics of the Czech Academy of Sciences$,$ Na Slovance 2$,$ 18221 Praha 8$,$ Czech Republic}%

\author{Jirka Hlinka}%
\affiliation{ 
FZU-Institute of Physics of the Czech Academy of Sciences$,$ Na Slovance 2$,$ 18221 Praha 8$,$ Czech Republic}%

\date{\today}

\begin{abstract}
Ferroelectric charged domain walls are known for their high electrical conductivity, making them promising candidates for modern electronics. 
A remarkably high conductivity and nominal charge density has been found in  the head-to-head ferroelastic domain wall of tetragonal barium titanate.
Conductivity of this domain wall decreases by several orders of magnitude when the temperature drops down below about 5 degrees Celsius when the tetragonal phase transforms to the orthorhombic one.
We  explored the evolution of these ferroelectric charged domain walls in BaTiO$_3$ crystals while they undergo this phase transition by in-situ optical microscopy.
Our results reveal that, below the phase transition, the domains adjacent to charge domain walls become twinned and the head-to-head charged domain wall transforms into a superdomain wall, which is broken into  alternating micron-scale segments with and without the excess bound charge.
 Since the macroscopic conductive channel along such fragmented superdomain wall is disrupted, these observations  explains the observed loss of the domain wall conductivity below the phase transition.
\end{abstract}

\maketitle

Charged domain walls (CDW) in ferroelectric materials are interfaces that can exhibit distinct electrical properties compared to the surrounding domains\cite{Meier2015functional,Sluka2016charged,bednyakov2018physics}. 
They can be created, displaced or erased inside a nominally isolating dielectric material by an electric field, which can be of interest for future electronic circuitry. 
The normal component of the spontaneous polarization has a jump at these interfaces, implying that their formation requires compensation by free charge ionic or electronic carriers \cite{tagantsev2010domains}.
CDWs can be either negatively or positively charged.
Negatively charged CDWs, or tail-to-tail polarization configurations, require positive screening charges (e.g., oxygen vacancies or holes), whereas positively charged CDWs, or head-to-head configurations, are stabilized by negative charges (e.g., electrons).

Previous research has demonstrated that positively charged CDWs in BaTiO$_3$ can exhibit a remarkable metallic-like conductivity up to 10$^9$ times higher than the bulk crystal \cite{sluka2013free,beccard2022nanoscale}.
This  conductivity has been attributed to a high density of charge carriers confined within the CDWs\cite{maksymovych2012tunable,sluka2013free}, suggesting the presence of a quasi-two-dimensional electron gas \cite{sluka2013free}.
Such two-dimensional electron systems are of particular interest due to their potential to display quantum phenomena, such as superconductivity and the quantum Hall effect.

However, it was also reported that, when BaTiO$_3$ was cooled from its tetragonal phase to the orthorhombic phase, the conductivity of the CDWs dramatically decreased \cite{sluka2013free}.
This observation rises an important question whether the wall and the compensating charge were only displaced from the electroded area or whether it is still a charged domain wall but with a much lower conductivity.

To address this question, we have revisited this temperature drop of the wall conductivity and explored the temperature evolution of head-to-head and tail-to-tail CDWs  in BaTiO$_3$ single crystal on cooling  through the phase transition from the tetragonal to the orthorhombic phase by means of polarized optical microscopy.
It turned out that macroscopic domain walls persist in the orthorhombic phase at the places of the former head-to-head and tail-to-tail tetragonal CDWs, but these new walls are fragmented in pieces.
We discuss the mechanism of this fragmentation and argue that the overall conductive path is necessarily interrupted by such inhomogeneities.

The $\langle 110 \rangle$ oriented barium titanate (BaTiO$_3$) single crystals were obtained through the top-seeded solution growth technique (TSSG) from Electro-Optics Technology GmbH.
The samples were shaped as bars with dimensions of 5x1x0.8\,mm
along the $[1\bar{1}0]$, $[001]$ and $[110]$ crystallographic directions, respectively.
The (001) plane was polished at ambient conditions to a quality of 1\textmu m, platinum electrodes were applied to the (110) planes and the domain structure was observed along the [001] direction.
Surface electrodes were contacted with a high-temperature silver paste.
Ferroelectric CDWs were created by poling the previously annealed sample along [110] direction while illuminating it with ultraviolet light source of 365\,nm wavelength and luminous flux about 5\,W, based on the light-emitting diode OSRAM LZ4-V4UV0R.

For temperature studies, the sample was placed in a Linkam cell allowing cooling down to the liquid nitrogen temperature (77\,K).
A  2\,kV/cm electric field was applied and the current along the [110] 
direction was measured while cooling the sample at the rate of about 1\,K/min using a Keithley 6517B electrometer with an integrated 1\,kV DC power supply.
In order to exclude the possibility that  charged domain wall is simply drifting from the electrode area,
we performed conductivity measurements with the  electrodes deposited across the entire sample, otherwise following a similar methodology as in Ref.\,\onlinecite{sluka2013free}.
The domain structure was examined using a polarizing microscope (Leica DM2700M) in transmission and reflection modes.
The observed domain patterns were documented using Leica's LAS X software.
All processes were automated within the LabVIEW environment.

\begin{figure}[t]
\centering
\includegraphics[width=.95\columnwidth]{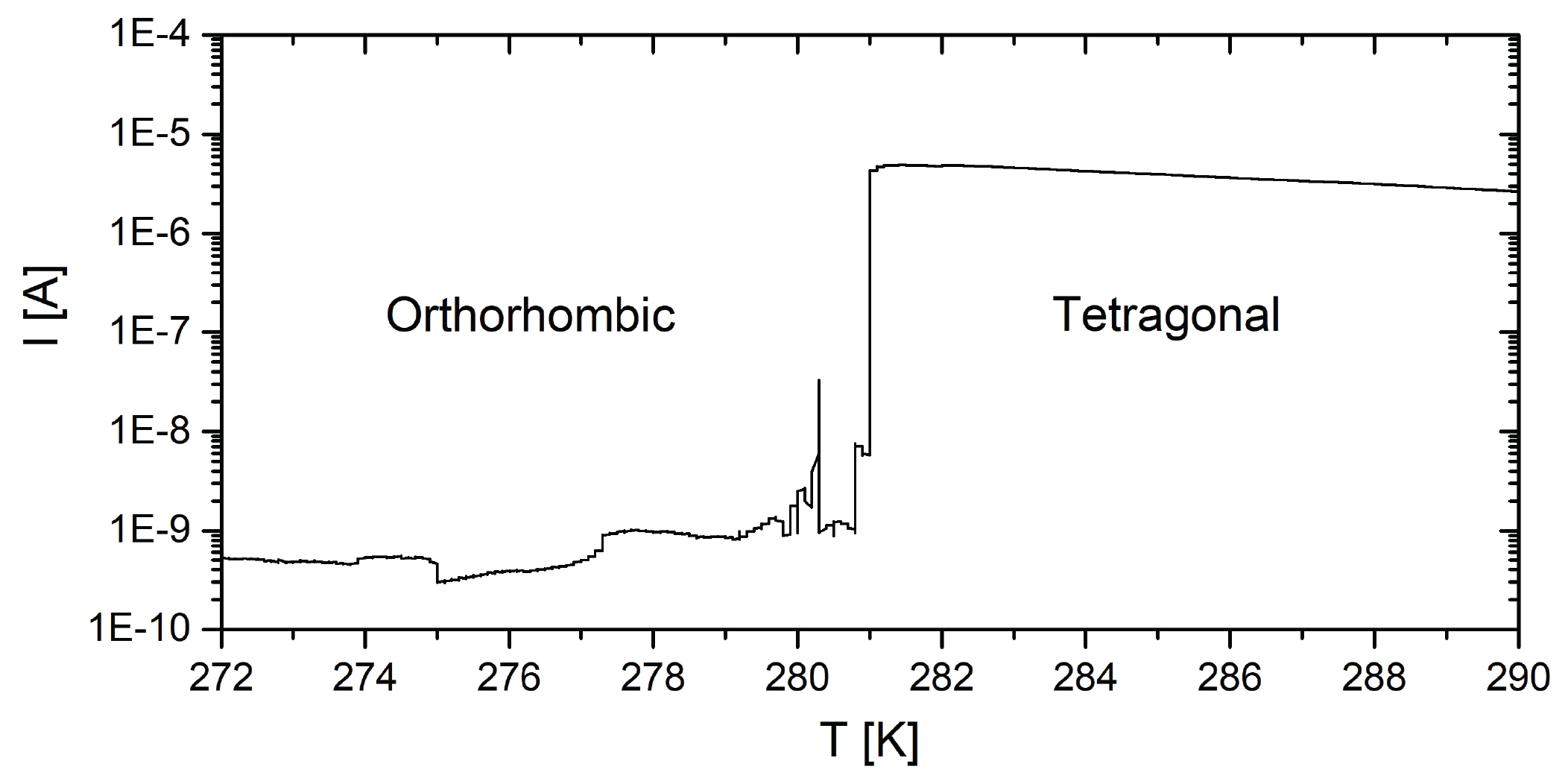}
\caption{
Current through the sample under the bias field of 2\,kV/cm on cooling  through the  phase transition to the orthorhombic phase, located near about 280\,K. The current flows along the [110] crystallographic direction between electrodes fully covering the 
(110) surfaces of the sample.
}
\label{Fig1}
\end{figure}

Resulting current is displayed in Fig.\,\ref{Fig1}.
Considering the reported conductivity values of head-to-head CDWs\,\cite{sluka2013free}, the effective conductivity of the entire sample in the tetragonal phase can be explained by their contributions only.
 The significant current drop by approximately four orders of magnitude observed in the orthorhombic phase (Fig.\,\ref{Fig1}) is also
 consistent with the findings of Ref.\,\onlinecite{sluka2013free}.

\begin{figure}[t]
\centering
\includegraphics[width=.9\columnwidth]{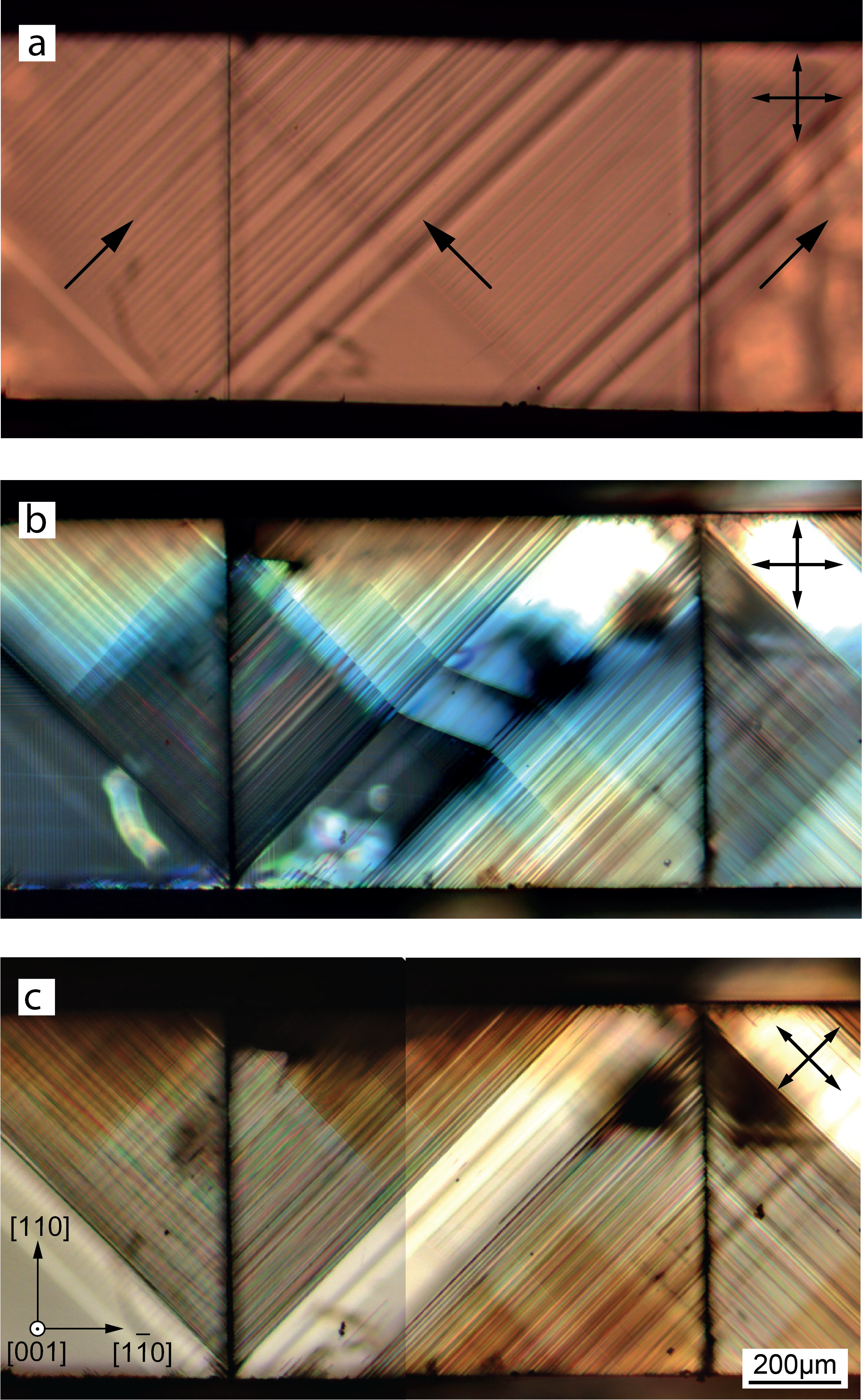}
\caption{
Observations of the poled sample by the polarizing microscope in a reflection mode. 
The crossed polarizer-analyzer orientation is shown by the crossed directors in the top-right corner of each panel, the black single-head arrows stands for the spontaneous ferroelectric polarization direction. 
Panel (a) shows one vertical head-to-head CDW and one vertical tail-to-tail CDW in the tetragonal phase at room temperature with polarizers in the bright orientation.
Panels (b) and (c) show the same area in the orthorhombic phase, with the same and 45$^{\circ}$ rotated position of crossed polarizers at 273K.
}
\label{Fig2}
\end{figure}

Simultaneously with the conductivity measurements, the domain structure was inspected.
Fig.\,\ref{Fig2}a shows the optical image recorded at the ambient conditions in the tetragonal phase.
The two vertical lines are caused by one head-to-head CDW and one  tail-to-tail CDW, separated by about 1\,mm thick monodomain area.
The inclined lines in the image are due to the inverse clapping angle surface ridges, created by a two-step process consisting in  polishing the sample in the presence of neutral ferroelastic domain walls existing in the pristine sample and in the subsequent removal of these domain walls during the poling procedure \cite{bednyakov2015formation}.
Orientation of polarization vectors in tetragonal domains, marked by black arrows in Fig.\,\ref{Fig2}a,  was determined considering the extinction position of the crossed polarizers,  the [110] poling field direction  and also the sense inverse clapping angles.

In the orthorhombic phase, the two vertical CDWs  persist at the same locations but they appear significantly broader in comparison with the narrow CDW of the tetragonal phase.
Moreover, in their vicinity, there is a twinned area with a new system of dense parallel  stripes, inclined at $\pm45^{\circ}$ with respect the vertical CDW lines (see Fig.\,\ref{Fig2}c).
In overall, this structure exhibits a high stability, with only minimal changes under an electric field varied up to 6 kV/cm.
How to understand these images?
In the center of the image, between the two twinned areas, there is a 200-micron thick homogeneous band. This band is  expected to be a  monodomain area with the preferred [110]-orientation, selected by the previous poling field and the in-situ applied voltage.
This is supported by the extinction position rotated by 45$^{\circ}$ with respect to that of the tetragonal phase. In particular, the monodomain area is dark in Fig.\,\ref{Fig2}b and bright in Fig.\,\ref{Fig2}c.
The same trend of the intensity variation with the crossed polarizer rotation is  observed in the twinned area. 
Therefore, the fine stripes 
have spontaneous polarization mostly along [110] and $[\bar{1}10]$ axes.
Moreover, to yield the average polarization along the poling field, the twinned area in the central part of Fig.\,\ref{Fig2}c between the two vertical CDWs must be formed by
the voltage-favored [110]-polarized stripe domains, alternated with $[\bar{1}10]$-polarized stripe domains.
Likewise, the twinned areas in the left and right side of Fig.\,\ref{Fig2}c  are formed by the voltage-favored [110]-polarized domains alternated with $[1\bar{1}0]$-polarized domains.

\begin{figure}[t]
\centering
\includegraphics{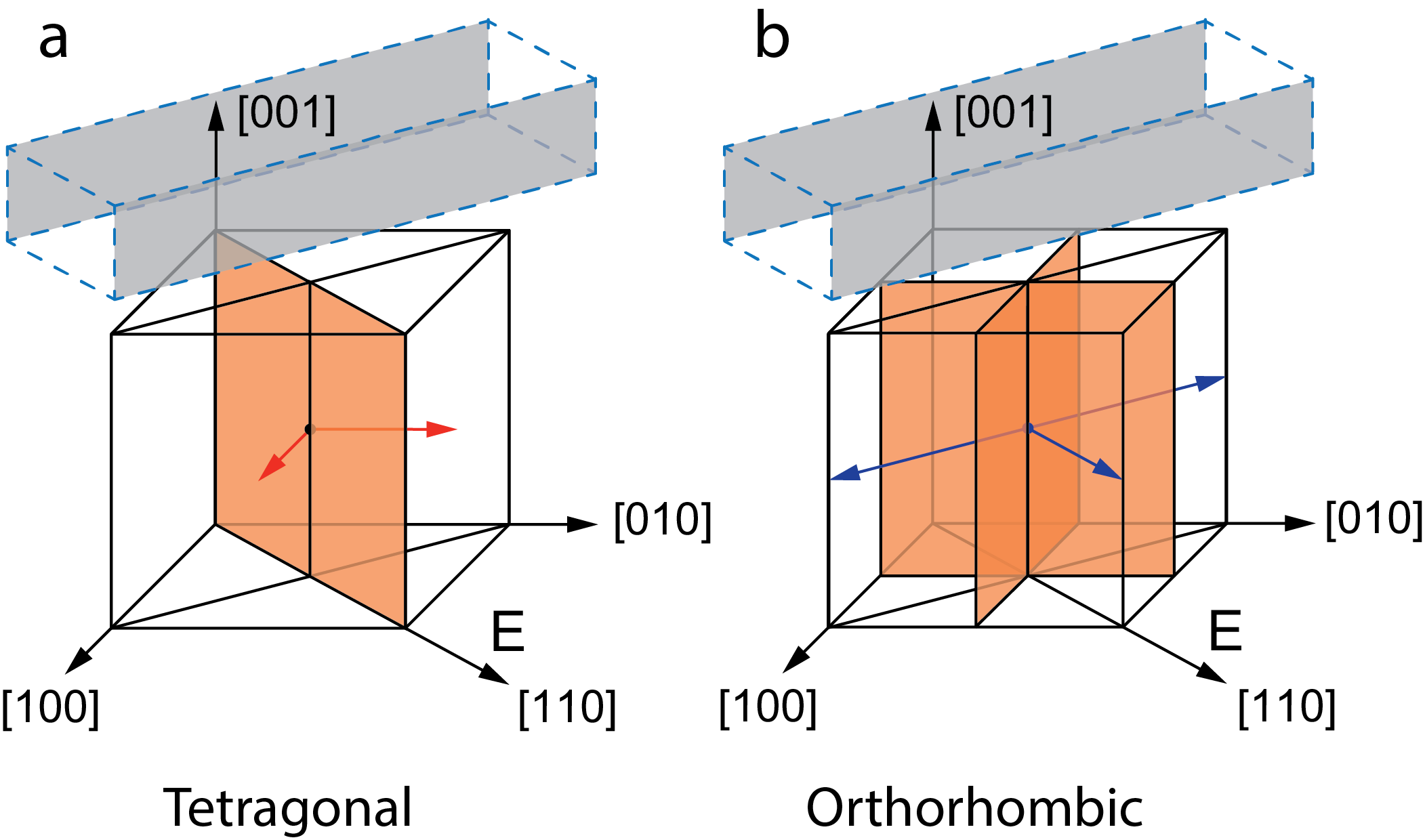}
\caption{
Orientation of the  mechanically compatible domain walls (orange planes) within [110] poled sample from the present observations in the  tetragonal (a) and orthorhombic (b) phases.
The sample shape is schematically shown in the top of each panel, the electrode surfaces are marked in gray.
Identified directions of the spontaneous ferroelectric polarization are indicated by red and blue arrows. The letter E stands for the [110] poling field direction.
}
\label{Fig3}
\end{figure}

The crystallographic orientation of the above identified domain states and domain walls in the present experiment is summarized in Fig.\,\ref{Fig3}. 
In the tetragonal phase, two domain states with polarization along [100] and [010] are present. 
The observed charged domain wall orientation ($\bar{1}10$) agrees with the mechanical compatibility condition for these two states. (Fig.\,\ref{Fig3}a).
In the orthorhombic phase, we have found one voltage-favored [110] domain state, which alternates with one of the two complementary $[\bar{1}10]$ and $[1\bar{1}0]$ domain states in the twinned areas.
Domain walls within the twinned areas are parallel to (100) and (010) planes and and they also fulfill the mechanical compatibility condition for these domain state pairs. (Fig.\,\ref{Fig3}b).
However, none of the above orthorhombic domain states form a pair satisfying the mechanical compatibility at the ($\bar{1}10$) plane.
Therefore, the planar interface we observed in the orthorhombic phase at about the same  position as where the CDW was formed in the tetragonal phase cannot be an ordinary domain wall.
Instead, it can be a superdomain wall,  a macroscopic boundary separating two finely twinned areas, and thus a boundary with an inherent heterogeneity at the scale of few microns.

\begin{figure}[t]
\centering
\includegraphics[width=.9\columnwidth]{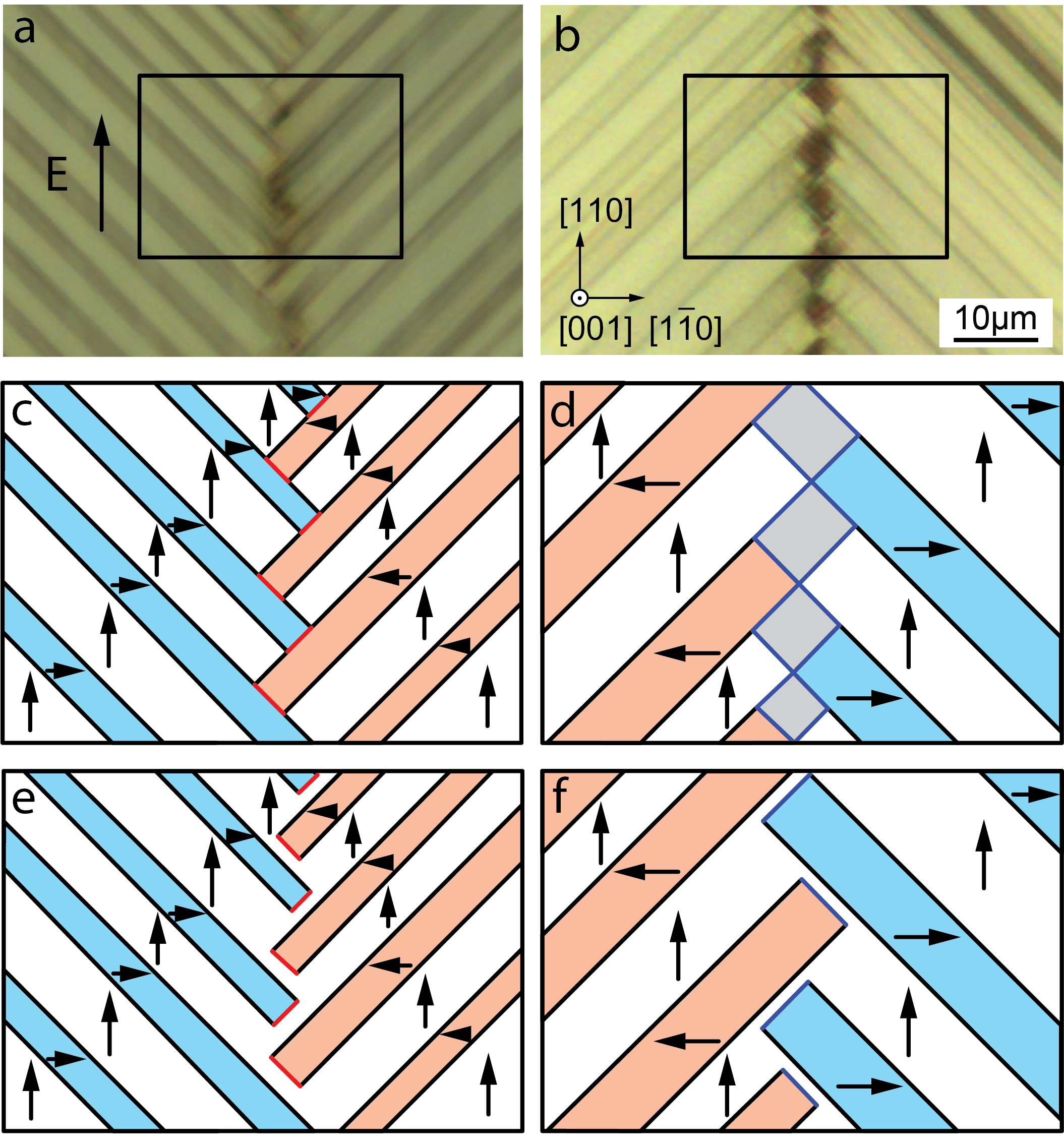}
\caption{
Micrographs of the domain structure in orthorhombic phase in the vicinity of head-to-head and tail-to-tail CDWs created in the tetragonal phase (a, b), their general schematic interpretation assuming interrupted stripes of the principal domain state (c, d) and percolated stripes of the principal domain state (e, f), respectively.
Regions with bound charge  marked as red segments in (c,e,f) and read rectangles in (d) may possibly have a zigzag or similar internal structure at the nanoscale.
}
\label{Fig4}
\end{figure}

The heterogeneity of head-to-head and tail-to-tail superdomain walls is apparent from the more magnified images  provided in Fig.\,\ref{Fig4}.
In these images the voltage-favored stripes have the same color and shade as the larger monodomain area (beyond the range of the Fig.\,\ref{Fig4}), while
the $[1\bar{1}0]$ and $[\bar{1}10]$-polarized stripe domains appear with a darker shade. 
The superdomain CDW is formed by zipping of primary ferroelastic twin stripe domains  and so it acquires thickness of about 3-5 microns, derived from the thickness of the individual stripes in the adjacent superdomains (see Fig.\,\ref{Fig4}a,b).


The existence of strong CDWs implies a build up of bound charge density that needs to be compensated by a free charge density of the opposite sign but a practically same magnitude.
In case of the head-to-head T90($1\bar{1}0$) CDW in the tetragonal barium titanate, the bound charge is positive and due to the 45$^{\circ}$ angle between the axes of the CDW normal and of the adjacent spontaneous polarization, its magnitude is $\sigma_{\rm T} = \sqrt{2} P_{\rm S}$, where $P_{\rm S}$ is the magnitude of the spontaneous polarization vector in the tetragonal phase.
It is reasonable to assume that the phase front of the structural order parameter in the orthorhombic phase propagates much faster than the effective macroscopic drift of the released free charge carriers.
Therefore, at least on the macroscopic scale, the orthorhombic polarization should accommodate to the same amount of the built-in free charge as it was needed for compensation of the original tetragonal CDWs.
This in turn implies  that just after passing the phase transition,  formation of CDW should be expected at the same planes and even the macroscopic bound charge distribution in the orthorhombic phase should be the same as it was in the tetragonal phase.

In the orthorhombic phase, however, the angle between the ($1\bar{1}0$) CDW axis and the spontaneous polarization axis in the adjacent domain can be either 0, 60 or 90$^{\circ}$, as opposed to 45$^{\circ}$ encountered for T90($1\bar{1}0$) CDW.
Assuming that the bound charge contribution from both sides of the CDW is the same in the orthorhombic phase, the corresponding net bound charge densities are $\sigma_{\rm O} = 0$,  $P_{\rm S}'$, and $2P_{\rm S}'$, respectively,  where $P_{\rm S}'$ is the magnitude of the spontaneous polarization in the orthorhombic phase.
Even if we assume that $P_{\rm S}'$ could be by 10-20 percent larger than $P_{\rm S}$, we still have 
$0 <  P_{\rm S}' < \sigma_{\rm T} < 2P_{\rm S}'$,
implying that none of the above primary orthorhombic CDW densities match the prescribed tetragonal $\sigma_{\rm T}$.
Thus, the bound charge density mismatch is just sufficiently large to explain the observed twinning near the orthorhombic charged superdomain walls. 

More precisely, in order to achieve the optimal average bound charge density, one of the primary domain states involved in the twin should have a larger bound charge density, and the other should have a smaller bound charge density than the aimed average  value $\sigma_{\rm T}$.
There is only one domain state capable to create a larger bound charge contribution  -- that with the polarization perpendicular and directed towards the  head-to-head domain wall with the considered orientation.
The inequalities $0 <  P_{\rm S}' < \sigma_{\rm T} < 2P_{\rm S}'$ imply that for the smaller bound charge state, one can speculate about the polarization at 60$^{\circ}$ with respect to the ($1\bar{1}0$) axis of the  wall.
The experimentally identified state with 0$^{\circ}$ is nevertheless the most natural choice because it has polarization in the direction of the applied voltage.
In addition, the pair of the experimentally identified states in the twins are mechanically compatible, connected by the lowest-energy neutral domain walls according to the Ginzburg-Landau-Devonshire calculations, the O90\{100\} walls\cite{marton2010domain}, and they allow to maintain the overall macroscopic mirror symmetry located at the CDW plane.
Last but not least, the average bound charge matching condition implies that the fraction of the voltage-favored primary domain state should be about
$\nu \doteq 0.7 P_{\rm S}/ P_{\rm S}'$.
Considering that $P_{\rm S}'$ can be by 10-20 percent\cite{wieder1955electrical, marton2010domain} larger than $P_{\rm S}$, we derive that this fraction is $\nu \approx 60$ percent, which is indeed in a very good rough agreement with the observations in the optical images.


The details of the images of head-to-head and tail-to-tail superdomain walls look qualitatively different,
suggesting that the role of qualitatively different charge carriers involved in the compensation of the bound charge is also manifesting in the nanoscale pattern formation.
In case of the head-to-head superdomain wall, the simplest interpretation of the observed image suggests that the stripes with polarization oriented perpendicularly to the superdomain wall are simply terminated on the stripes with the opposite polarization, forming thus few micron-wide charged 180$^{\circ}$ wall segments, highlighted in red in Fig.\,\ref{Fig4}c.
These are alternating with the regular, mechanically compatible 90$^{\circ}$  neutral domain wall segments (see  Fig.\,\ref{Fig4}c).
The existence of the charged 180$^{\circ}$ wall segments suggests that a considerable amount of compensating electronic charge carriers are present at the same place.
On the other hand,  the 90$^{\circ}$  neutral domain wall segments should not attract these charge carriers and so  a rather low conductivity is expected there.
In this way the conducting channel is interrupted and overall current is expected to be drastically reduced.

Another possibility is that the dominant domain state with polarization along the poling field is percolated across the superdomain wall,
as shown in Fig.\,\ref{Fig4}e.
In this case the stripes with polarization oriented perpendicularly to the superdomain wall can be terminated on the dominant
state by a mechanically compatible 90$^{\circ}$  charged domain wall segment, possibly collecting the carriers and having a large
conductivity.
The conducting channel is also interrupted there, as clear from Fig.\,\ref{Fig4}e.

In case of the tail-to-tail superdomain wall, the corresponding plausible interpretation with  mechanically compatible walls is given in Fig.\,\ref{Fig4}d,f.
The voltage-favored  stripes and 
the stripes with polarization perpendicular to the superdomain wall in Fig.\,\ref{Fig4}d seems to be terminated by similar dark patterns.
It suggests a more regular distribution of the compensating charge along the tail-to-tail superdomain wall. 
Due to the nature of the compensating charge carriers, however, the high conductivity is not expected to occur at the tail-to-tail superdomain wall at all.

Our optical observation  does not allow to conclude about the  submicron structure of the superdomain wall. It is likely that the dark areas within superdomain walls shown in Figs. \ref{Fig4}a,b are actually made by nanoscale wedges or zig-zag junctions \cite{ignatans2020local} but this does not disqualify our analysis. 

In summary, the present study reveals how the head-to-head and tail-to-tail charged domain walls created in tetragonal barium titanate are spontaneously replaced by a similarly oriented and located charged interfaces in the orthorhombic phase when the sample is cooled down to the temperatures below about 5 degrees Celsius.
These new boundaries have character of  heterogeneous superdomain walls and they imply a fine twinning in the neighbourhood of these boundaries.
Superdomain and superdomain wall formation can be explained as a result of conservation of the built-in free charge and  mechanical compatibility at the coarse scale.
Our analysis clarify that the drop down of the remarkable conductance of the head-to-head CDW below the tetragonal to orthorhombic phase transition is caused by breaking the continuity of the free-charge distribution along the superdomain wall.
These findings are likely to help in investigation of the peculiar properties of barium titanate CDWs.

The authors acknowledge the support from the Czech Science Foundation (GACR project No.~20-05167Y) and the assistance provided by the Ferroic Multifunctionalities project, supported by the Ministry of Education, Youth, and Sports of the Czech Republic. Project No. CZ.02.01.01/00/22\underline{\,\,\,}008/0004591, co-funded by the European Union.

\section*{Conflict of Interest}

The authors have no conflicts to disclose.

\section*{Author contributions}

P.S.B. and J.H. conceived the project and P.S.B. conducted the experiment. 
P.S.B and J.H. were responsible for writing the manuscript.
I.R. provided technical support and contributed to the preparation of samples.
All authors contributed to the development and discussions of the ideas presented in the paper, revised and approved the final version of the manuscript.

\section*{Data Availability Statement}

All data generated or analyzed during this study are included in this article and available from the authors upon reasonable request.

\bibliography{References}

\end{document}